\documentclass[11pt,a4paper]{article}
\usepackage{jheppub}
\usepackage{epsfig}
\usepackage{amsfonts}
\usepackage{amscd}
\usepackage{latexsym}
\usepackage{amsmath,amssymb}
\usepackage{verbatim}
\usepackage{setspace}  
\usepackage{hyperref}
\usepackage{slashed}
\usepackage{float}
\usepackage{physics}
\usepackage{marvosym}
\usepackage[percent]{overpic}
\usepackage[compact]{titlesec}
\usepackage{soul}

\numberwithin{equation}{section}
\def\p{\partial}
\def\kap{\kappa}
\def\ca{{\mathcal A}}

\def\co{{\mathcal O}}

\def\cq{{\mathcal Q}}

\def\<{\langle }
\def\>{\rangle}
\def\eps{\varepsilon}

\def\Lam{\Lambda}
\def\half{\frac{1}{2}}

\newcommand{\ba}{\begin{aligned}}
\newcommand{\ea}{\end{aligned}}
\def\beq{\be\begin{array}{c}}
\def\eeq{\end{array}\ee} 
\def\be{\begin{equation}}
\def\ee{\end{equation}}
\def\bea{\begin{eqnarray}}
\def\eea{\end{eqnarray}}

\title{Global First Laws of Accelerating Black Holes}
 
\author{Adam Ball}

\affiliation{Center for the Fundamental Laws of Nature, Harvard University,\\
17 Oxford Street, Cambridge, MA 02138, USA}

\emailAdd{aaball@g.harvard.edu}

\abstract{We generalize the first law of black hole mechanics to the rotating, charged C-metric and to the Ernst metric, both of which have the charged C-metric as a special case. All of these metrics are (3+1)-dimensional, have vanishing cosmological constant, and physically describe a pair of black holes pulled apart to null infinity by some external force. Our first laws are global in the sense of applying to an entire patch of spacetime, as opposed to a neighborhood of the black hole. They are formulated with respect to ``boost time", whose primacy is motivated by the celestial holographic approach to scattering amplitudes.}

\begin{document} 
\maketitle
\flushbottom

\section{Introduction}

Despite the substantial importance of and interest in black holes, there is a striking paucity of exact black hole solutions in general relativity, due in part to the no-hair theorem \cite{Chrusciel:2012jk}. In particular, in 3+1 dimensions and with vanishing cosmological constant there are only a few known generalizations of the Kerr-Newman metric. One is the family of boost-rotation symmetric black hole spacetimes \cite{pravda:2000vh}, which physically describe a pair of uniformly accelerating black holes pulled apart to null infinity by some external force. They exhibit a stunning variety of phenomena, including but by no means limited to generalized Rindler physics, black hole pair production \cite{Garfinkle:1990eq, Garfinkle:1993xk}, ER=EPR \cite{Garfinkle:1993xk, Maldacena:2013xja}, and superrotation vacuum transitions \cite{stromzhib}. Accelerating black holes clearly constitute a rich sandbox to test new ideas. This paper hands the reader a shovel by generalizing the first law of \cite{Ball:2020vzo} to the rotating, charged C-metric and to the Ernst metric, both of which have the charged C-metric as a special case. The work \cite{Ball:2020vzo} in turn improves on the result of \cite{Dutta:2005iy}. Our approach to first laws for accelerating black holes differs from the previous works \cite{Gibbons:2013dna, acovErnst, Astorino, AstReg, Anabalon} in that we take a more global perspective, incorporating the acceleration horizon, the cosmic strings, and spatial infinity, as opposed to a local perspective concerned only with a neighborhood of the black hole. See \cite{Ball:2020vzo} for further comparison of these perspectives. For the more tractable asymptotically AdS case see \cite{Appels:2017xoe, Anabalon:2018ydc, Anabalon}. We note also that while these metrics are genuine exact solutions, one can view any cosmic strings they contain as idealizations of finite-width strings that exist in gauge theories such as the Standard Model \cite{Garfinkle:1985hr, Achucarro:1995nu, Gregory:1995hd}.

Our first laws are formulated in terms of ``boost time", a generalization of Rindler time. This is motivated in large part by a connection to celestial holography. One usually formulates scattering amplitudes in terms of momentum eigenstates, but in celestial holography one instead scatters boost eigenstates, and reinterprets the amplitudes as the correlation functions of a ``celestial CFT" in two fewer dimensions \cite{Pasterski:2016qvg}. In Minkowski space boosts and rotations in a given direction are dual to dilations and rotations around a given point in the putative celestial CFT, and so in celestial holography they are of primary importance among the symmetries present. Boosts and rotations along the string axis are precisely the symmetries preserved by the accelerating black hole metrics we study, and so it seems natural that these black holes should have some role to play in celestial holography, whether that be classical backgrounds for scattering, coherent states in the quantum theory, or something else entirely \cite{stromzhib}. As exciting and successful as the celestial program is, nearly all work so far has been limited to perturbative analysis around flat space (in fact, most at tree level). Thus understanding the boost charges and semiclassical thermodynamics of these black holes offers a possible hint at non-perturbative aspects of celestial holography.

The structure of the paper is as follows. The first few sections focus on the rotating, electrically charged C-metric. In section \ref{sec:SCrev} we review the rotating C-metric and some of its basic properties. In particular it generalizes Rindler space and contains a pair of black holes with nonzero electric charge and angular momentum. These black holes are pulled apart to null infinity by a pair of cosmic strings. In section \ref{sec:cosmicback} we discuss the cosmic string background that is naturally associated with a given rotating C-metric and present a set of coordinates in which the rotating C-metric agrees with its background near spatial infinity at leading and first subleading order in a radial coordinate. In section \ref{sec:SCquant} we compute various quantities in the rotating C-metric that will show up in the first law. In section \ref{sec:SCFirst} we define and compute the boost mass of the patch of spacetime containing the black hole, and subsequently present the first law for the rotating C-metric. The boost mass turns out to vanish. The next three sections are devoted to the Ernst metric. In section \ref{sec:ErnstRev} we review the electrically charged Ernst metric with cosmic strings. It generalizes Rindler space and contains a pair of black holes with nonzero electric charge that are pulled apart to null infinity by a background electric field and cosmic strings. In section \ref{sec:Melv} we review the electric Melvin spacetime, which naturally serves as a background for the Ernst metric. We also present coordinates for the Ernst metric in which spatial infinity is more easily described. In section \ref{sec:Ernstqnf} we compute various quantities in the Ernst metric and present its first law. Its boost mass does not vanish, but can be understood in terms of an electric dipole moment. In section \ref{sec:conc} we conclude. We work in geometrized units, $G = c = 1$, throughout.

\section{Rotating C-metric review} \label{sec:SCrev}

The rotating, electrically charged C-metric (henceforth rotating C-metric), sometimes called the accelerating Kerr-Newman black hole, is a member of the Plebanski-Demianski family of boost-rotation symmetric Type D exact solutions \cite{Plebanski:1976gy}. Physically it describes two rotating, charged black holes pulled apart to null infinity by a pair of cosmic strings \cite{Hong:2004dm}. The rotating C-metric and related spacetimes have been discussed extensively, for example in \cite{Weyl, kinwalk, Plebanski:1976gy, hawkross, Hong:2004dm, Griffiths:2005se, Griffiths:2005mi, gkp, Astorino, Podolsky:2021zwr}. See \cite{gkp} for some illuminating Penrose diagrams. See also \cite{Ball:2020vzo} for a parallel discussion of the non-rotating, charged C-metric. The rotating C-metric has two Killing vectors, which are often referred to as the boost and rotation generators. The Killing horizons of the boost generator naturally divide the spacetime into patches. Our patch of interest contains one of the two black holes. This patch has access to half of spatial infinity and is partially bounded by a so-called acceleration horizon, which is a non-compact Killing horizon of the boost generator \cite{Dutta:2005iy}. The boost generator is timelike in this patch, and we refer to its natural coordinate as boost time. The metric of the rotating C-metric can be written as
\be \ba \label{SCm} ds^2 = \frac{1}{A^2 (x-y)^2} \Bigg( \frac{G(y)}{1 + (aAxy)^2} \left( \frac{1 + a^2 A^2 x^2}{1 + a^2 A^2} dt + \alpha a A (1 - x^2) d\phi \right)^2 - \frac{1 + (aAxy)^2}{G(y)} dy^2 \\
+ \frac{1 + (aAxy)^2}{G(x)} dx^2 + \frac{G(x)}{1 + (aAxy)^2} \left( \alpha (1 + a^2 A^2 y^2) d\phi - \frac{aA(1-y^2)}{1+a^2A^2} dt \right)^2 \Bigg). \ea \ee
This form is related to the coordinates of \cite{Hong:2004dm} by a linear transformation in $t$ and $\phi$.\footnote{Historically there was a similar but physically distinct member of the Plebanski-Demianski family that bore the same name of ``rotating" or ``spinning" C-metric. We use the term in the modern sense introduced in \cite{Hong:2004dm}.} It is an exact solution of the Einstein-Maxwell-Nambu-Goto equations, with cosmic struts/strings along the $\phi$-axis and gauge field given by
\be A_\mu dx^\mu = \frac{qy}{1 + (aAxy)^2} \left( \frac{1 + a^2 A^2 x^2}{1 + a^2 A^2} dt + \alpha a A (1 - x^2) d\phi \right). \ee
The function $G(\zeta)$ is the quartic polynomial
\be G(\zeta) = (1-\zeta^2)(1+Ar_+ \zeta)(1+Ar_-\zeta) \ee
with
\be r_\pm = m \pm \sqrt{m^2 - q^2 - a^2}. \ee
We assume $Ar_+ < 1$, so the four roots of $G(\zeta)$ satisfy
\be \frac{-1}{Ar_-} < \frac{-1}{Ar_+} < -1 < 1. \ee
\begin{figure}
  \centering
  \begin{minipage}[b]{0.49\textwidth}
    \centering
      \begin{overpic}[width=\textwidth]{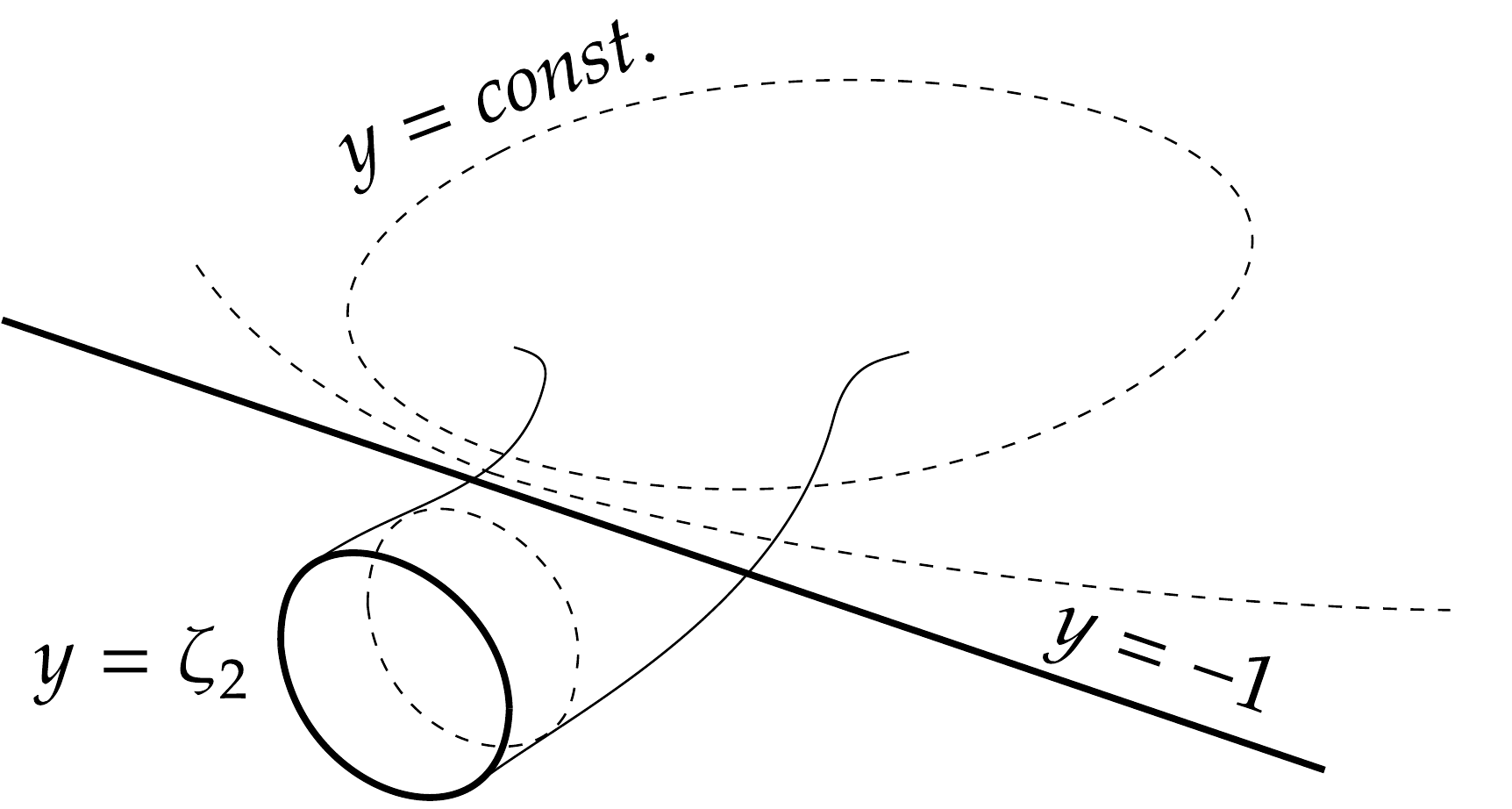}
        \put(9.5,3.8){\includegraphics[width=.085\textwidth]{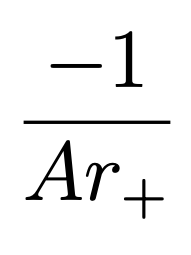}}
      \end{overpic}
    \caption{\label{fig:wormhole_y} Lines of constant $y$ on the $t = \rm const.$, $\phi = 0, \pi$ slice.}
  \end{minipage}
  \hfill
  \begin{minipage}[b]{0.49\textwidth}
  \centering
        \includegraphics[width=\textwidth]{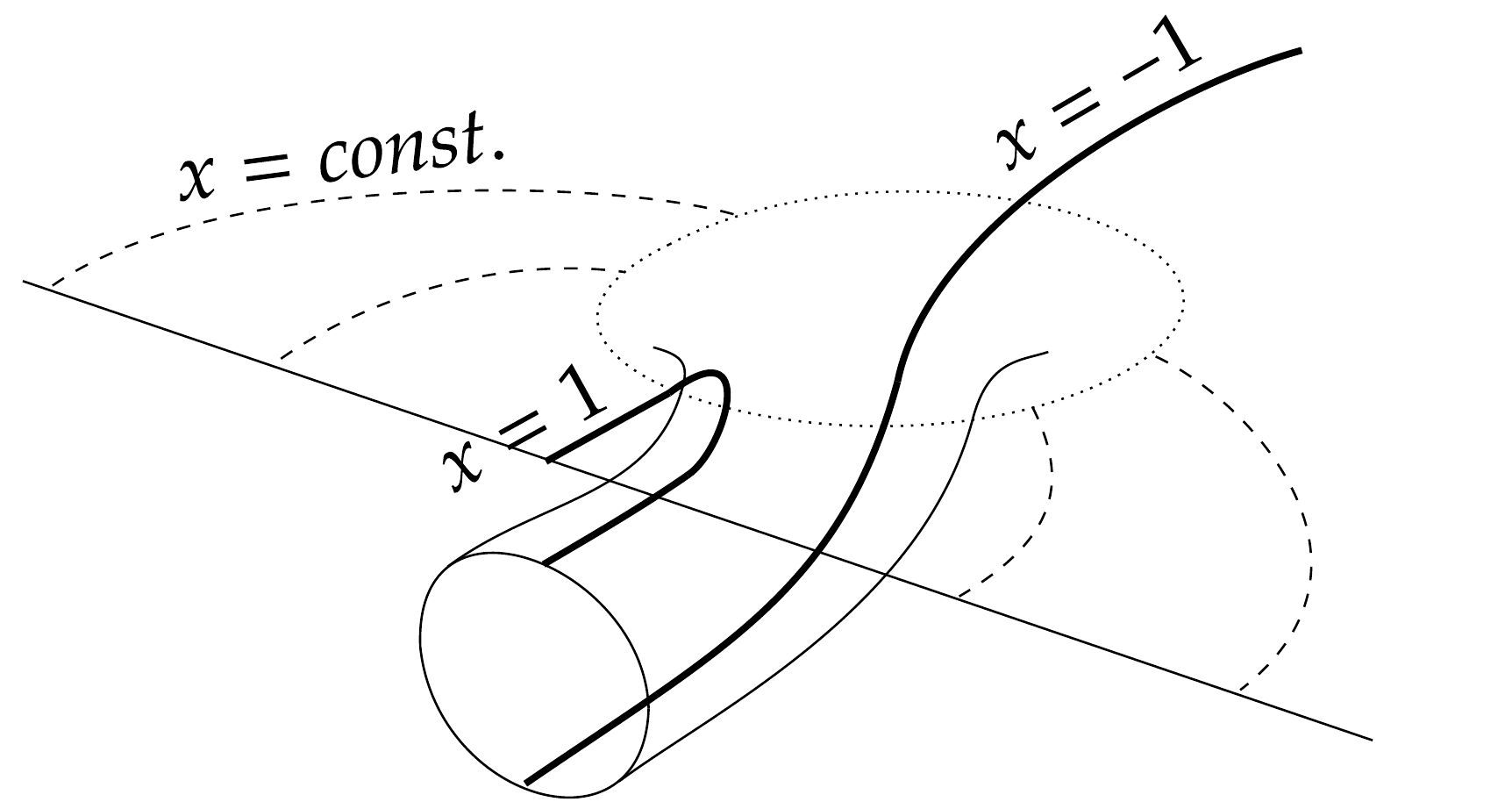}
    \caption{\label{fig:wormhole_x} Lines of constant $x$ on the $t = \rm const.$, $\phi = 0, \pi$ slice.}
  \end{minipage}
\end{figure}
\hspace{-1.3mm}We identify $\phi \sim \phi + 2\pi$ with other coordinates fixed. The range of $x$ is ${-1 \le x \le 1}$. The $\phi$-axis between the black holes corresponds to $x=1$, while the outer $\phi$-axis corresponds to $x=-1$. See figures \ref{fig:wormhole_y} and \ref{fig:wormhole_x}, reproduced from \cite{Ball:2020vzo}. We will concentrate on the patch with $\frac{-1}{Ar_+} \le y \le -1$, in which the boost generator $\p_t$ is timelike. As we will see, this patch is analogous to Rindler space. The acceleration horizon lies at $y=-1$ and the black hole horizon lies at $y = \frac{-1}{Ar_+}$. If we continued past the first black hole horizon we would encounter its second, inner horizon at $y = \frac{-1}{Ar_-}$. Spatial infinity is reached by sending $y, x \to -1$ simultaneously with $\frac{1+x}{-1-y}$ fixed.

A particular solution is characterized by five parameters, such as $A, m, q, a, \alpha$. Roughly speaking, the first four respectively give the black hole acceleration, mass, electric charge, and specific angular momentum. The parameter $\alpha$ could be absorbed into the range of $\phi$, but we prefer to keep $0 < \phi < 2\pi$. One may set $\alpha$ to eliminate the conical singularity between the black hole and acceleration horizon, but here we leave it free. Generically there will be a strut at $x=1$ between the black hole and acceleration horizon (we will always call it a strut, even though its deficit angle can have either sign), in addition to the usual string at $x=-1$ stretching from the black hole out to infinity.

To gain some intuition for the spacetime it helps to know some limits in which it simplifies. Defining $t' \equiv t/A, \, r \equiv \frac{-1}{Ay}, \, \theta \equiv \cos^{-1} x$ and sending $A \to 0, \alpha \to 1$ with $m, q, a$ fixed, we arrive at the Kerr-Newman metric in Boyer-Lindquist coordinates with mass $m$, charge $q$, and specific angular momentum $a$. This justifies the aforementioned approximate physical interpretations of $m$, $q$, and $a$. If we instead send $m, q, a \to 0$ with $A, \alpha$ fixed, we get a (locally) flat cosmic string spacetime, with conical deficit angle $\delta = 2\pi (1-\alpha)$. Further sending $\alpha \to 1$ yields Minkowski space, in coordinates such that $\p_t$ is the canonical boost generator in the direction of the $\phi$-axis. In this case the trajectory of the would-be black hole at $y \to -\infty$ is that of a point particle particle with uniform acceleration $A$, i.e. a Rindler particle.

Of all the special cases of the rotating C-metric, the connection to Rindler space is the most important to this paper. As explained in the introduction, this work aims to understand accelerating black holes from the point of view of boost time. Rindler time is future-directed only within a certain patch of Minkowski space, bounded by the acceleration horizon. This sense of time may seem strange, but that is no obstruction to computing physical phenomena seen by a Rindler observer such as the Unruh effect \cite{Fulling:1972md, Davies:1974th, Unruh:1976db}. Similarly our boost time $t$ is future-directed within $\frac{-1}{Ar_+} < y < -1$. It is within this patch that we envision our observers and compute our first law. The connection with Rindler space only deepens when we consider the asymptotic properties of the rotating C-metric. In particular, $\p_t$ is asymptotically proportional to a Rindler time translation near spatial infinity.

At this point we have more or less reviewed the setup in \cite{Ball:2020vzo}. The present generalization to include angular momentum leads to only a few qualitative differences. Although the black hole now carries angular momentum, the strut and string still do not. Accordingly, there are no closed timelike curves in this spacetime \cite{Hong:2004dm}. Still, the presence of a nonzero $g_{t\phi}$ metric component potentially complicates the periodicity of $\phi$. However, in our coordinates we have $\phi \sim \phi + 2\pi$ with no need for a simultaneous shift of $t$. A priori there is frame dragging near the horizons relative to infinity, and corresponding angular potentials which will show up in our first law.

\section{Cosmic string background} \label{sec:cosmicback}

The rotating C-metric is asymptotically locally flat, meaning that near spatial infinity it looks like the flat cosmic string spacetime \cite{Griffiths:2005se}. Thus it is natural to use the cosmic string spacetime as a background for the rotating C-metric, just as Minkowski space naturally serves as a background for asymptotically flat spacetimes. Typically a spacetime's charges are defined relative to its background. This will serve to regularize na\"ively divergent charges of the rotating C-metric later on. We record here coordinates for the rotating C-metric and the cosmic string background such that the metric components agree asymptotically near spatial infinity at leading and first subleading order in a radial coordinate. This subleading matching is a special property of the rotating C-metric, and does not hold for generic asymptotically locally flat spacetimes.

Despite the many virtues of the rotating C-metric coordinates in \eqref{SCm}, they are not convenient for describing spatial infinity, which is reached by sending $y, x \to -1$ with $\frac{1+x}{-1-y}$ fixed. A more apt set of coordinates is $(t,\eps,\theta,\phi)$, defined by
\be \label{yeps}  y = -1 - \eps \cos^2\theta \left( 1 - \eps \left[ \frac{(3\cos2\theta - 1) G''(-1)}{4G'(-1)} + \frac{2a^2A^2(2\cos2\theta - 1)}{1+a^2A^2} \right] \right), \ee
\be \label{xeps} x = -1 + \eps \sin^2\theta \left( 1 - \eps \left[ \frac{(3\cos2\theta + 1) G''(-1)}{4G'(-1)} + \frac{2a^2A^2(2\cos2\theta + 1)}{1 + a^2A^2} \right] \right). \ee
Now spatial infinity is reached by $\eps \to 0$, with $\theta, \phi$ parametrizing the direction. The full rotating C-metric in these coordinates is straightforward to obtain, but quite lengthy. Expanding in $\eps$ gives
\be \label{SCexpansion} ds^2 = \frac{1+a^2A^2}{A^2 G'(-1) \eps} \left( -\frac{G'(-1)^2 \cos^2\theta}{(1+a^2A^2)^2} dt^2 + \frac{d\eps^2}{\eps^2} + 4d\theta^2 + \alpha^2 G'(-1)^2 \sin^2\theta d\phi^2 \right) + \co(\eps), \ee
\be \label{SCAexp} A_\mu dx^\mu = \frac{-q}{1+a^2A^2} dt + \co(\eps), \ee
where $d\eps$ is considered $\co(\eps)$ and $dt, d\theta, d\phi$ are $\co(\eps^0)$. Aside from $d\eps d\theta$, the off-diagonal components in \eqref{SCexpansion} are identically zero. The leading terms shown in \eqref{SCexpansion} and \eqref{SCAexp} turn out to satisfy the equation of motion on their own, and we use them to define our background fields. Explicitly,
\be \label{bkgd} ds^2_{\rm bkgd} \equiv \frac{1+a^2A^2}{A^2 G'(-1) \eps} \left( -\frac{G'(-1)^2 \cos^2\theta}{(1+a^2A^2)^2} dt^2 + \frac{d\eps^2}{\eps^2} + 4d\theta^2 + \alpha^2 G'(-1)^2 \sin^2\theta d\phi^2 \right), \ee
\be \label{bkga} A^{\rm bkgd}_\mu dx^\mu \equiv \frac{-q}{1+a^2A^2} dt. \ee
This is just the flat cosmic string in unfamiliar coordinates, which can be seen by transforming to cylindrical coordinates $(T, Z, \rho, \phi)$ via
\be \ba \tanh \left( \frac{G'(-1)}{2(1+a^2A^2)} t \right) & = T/Z, \\
\frac{4(1+a^2A^2)}{A^2G'(-1)\eps} & = -T^2 + Z^2 + \rho^2, \\
\tan \theta & = \frac{\rho}{\sqrt{Z^2 - T^2}}, \ea \ee
which yields
\be ds^2_{\rm bkgd} = -dT^2 + dZ^2 + d\rho^2 + \alpha^2 \frac{G'(-1)^2}{4} \rho^2 d\phi^2. \ee
This metric is flat everywhere except along the axis $\rho = 0$, where there is a conical singularity of deficit angle
\be \delta = 2\pi \left( 1 - \half \alpha G'(-1) \right). \ee
We see in \eqref{SCexpansion} that the rotating C-metric differs from its background only at sub-subleading order in $\eps$. Equivalently, viewing the rotating C-metric as a perturbation $h_{\mu\nu}$ on the background $g_{\mu\nu}^{\rm bkgd}$, in an orthonormal frame its components are $h_{ab} = \co(\eps^2)$. The electromagnetic field strength falls off similarly. In an orthonormal frame one finds $F_{ab} = \co(\eps^2)$.

Surfaces of constant $t$ and small $\eps$ are approximately large hemispheres of radius $\eps^{-1/2}$, with polar angle $\theta$ and azimuthal angle $\phi$. They are hemispheres rather than spheres because at $\theta = \frac{\pi}{2}$ we encounter the acceleration horizon. When comparing charges in the rotating C-metric and the cosmic string background one usually regulates the spacetimes with large but finite boundaries near their respective spatial infinities such that these boundaries have the same intrinsic geometry \cite{hawkhor}. We instead use surfaces of fixed $t, \eps$ with the same $\eps$ for both the rotating C-metric and the cosmic string background. This introduces an error at sub-subleading order in $\eps$. Clearly this does not affect charges that are already finite as $\eps \to 0$. Conveniently, it does not affect our regularized charges either since (one can show) the divergence is only $\co(\eps^{-1})$, so that the finite piece is at first subleading order and suffers no error from our approximation.

\section{Rotating C-metric physical quantities} \label{sec:SCquant}

In this section we record some physical quantities in the rotating C-metric in our conventions. All can be viewed as functions of $A, m, q, a, \alpha$. The area of the black hole is
\be \ca_{\rm bh} = \frac{4\pi\alpha(r_+^2 + a^2)}{1 - A^2 r_+^2}. \ee
The change in the acceleration horizon area relative to the background is straightforward to compute using \eqref{SCm}, \eqref{xeps}, and \eqref{bkgd}. See the analogous case in \cite{Ball:2020vzo} for more details. One finds
\be \Delta \ca_{\rm acc} = -\frac{4\pi\alpha (m(1 - A^2a^2) - Aq^2)}{A(1 - Ar_+)(1 - Ar_-)}. \ee
The electric charge of the black hole is
\be Q = \frac{1}{4\pi} \oint dx d\phi \sqrt{-g} \, F^{yt} \big|_{y=\frac{-1}{Ar_+}} = \alpha q. \ee
The acceleration horizon is also associated with an electric charge,
\be Q_{\rm acc} = \frac{1}{4\pi} \oint dx d\phi \sqrt{-g} \, F^{yt} \big|_{y=-1} = \alpha q. \ee
Since $Q_{\rm acc} = Q$ we will only use $Q$ in what follows. The black hole's angular momentum can be computed as the Komar charge for $\p_\phi$ or equivalently with covariant phase space methods, to be introduced in section \ref{sec:SCFirst}. Integrating over the black hole horizon one finds
\be J = \alpha^2 a m. \ee
The acceleration horizon has an angular momentum as well,
\be J_{\rm acc} = \alpha^2 a m, \ee
and we note $J_{\rm acc} = J$. Generally the tension of a cosmic string is related to its deficit angle by $\delta = 8\pi \mu$. The tensions of the strut at $x=1$ and string at $x=-1$ are
\be \mu_{\rm strut} = \frac{1}{4} \left( 1 - \half \alpha |G'(1)| \right), \ee
\be \mu_{\rm string} = \frac{1}{4} \left( 1 - \half \alpha G'(-1) \right). \ee
Several of our quantities of interest are defined in terms of Killing vectors. We will take $\xi = N \p_t$ as our sense of boost time, with normalization $N(A, m, q, a, \alpha)$. We use this same normalization for the Killing generators of the horizons. The black hole horizon has Killing generator
\be \chi_{\rm bh} = \xi + \Omega_{\rm bh} \p_\phi \ee
with
\be \Omega_{\rm bh} = -\frac{N a (1 - A^2 r_+^2)}{\alpha A (1 + a^2 A^2)(r_+^2 + a^2)}. \ee
The acceleration horizon has
\be \chi_{\rm acc} = \xi + \Omega_{\rm acc} \p_\phi \ee
with
\be \Omega_{\rm acc} = 0. \ee
In a different rotation frame $\Omega_{\rm acc}$ would be nonzero, so we keep it explicit. The surface gravity of the black hole horizon with respect to $\chi_{\rm bh}$ is
\be \kap_{\rm bh} = \frac{N |G'(\frac{-1}{Ar_+})|}{2(1+a^2/r_+^2)}. \ee
The surface gravity of the acceleration horizon with respect to $\chi_{\rm acc}$ is
\be \kap_{\rm acc} = \frac{N G'(-1)}{2(1+a^2A^2)}. \ee
The electric potential at the black hole horizon is
\be \Phi_{\rm bh} \equiv (\chi_{\rm bh}^\mu A_\mu)|_{y = \frac{-1}{Ar_+}} = -\frac{N q \, r_+}{A(r_+^2 + a^2)}. \ee
Note this is computed with the Killing generator $\chi_{\rm bh}$, not with the boost time $\xi$. The same point holds for the standard Kerr-Newman case. The electric potential at the acceleration horizon is
\be \Phi_{\rm acc} \equiv (\chi_{\rm acc}^\mu A_\mu)|_{y = -1} = -\frac{N q}{1 + a^2A^2}. \ee
We use a generalized sense of length for the cosmic strut and string. The usual sense of length of a string or other spatially 1D object in flat space has multiple generalizations to curved space. The most salient one is perhaps fixing a time slice and using the proper length of the string within that slice. Another approach is to define the length as the string's worldsheet area per unit time \cite{Herdeiro:2009vd, Krtous:2019fpo}. This latter sense turns out to be appropriate in our case. We find for the strut that
\be \ell_{\rm strut} \equiv \int_{-1/Ar_+}^{-1} dy \, \xi^t \sqrt{-g_{tt} g_{yy}} \Big|_{x=1} = \frac{N(1 - A r_+)}{2A^2 (1 + A r_+).} \ee
The change in the generalized length of the string relative to the background can be computed using \eqref{SCm}, \eqref{yeps}, and \eqref{bkgd}. See the analogous case in \cite{Ball:2020vzo} for more details. One finds
\be \Delta \ell_{\rm string} = -N \frac{1 - 3a^2 A^2 - Ar_- (3 - a^2 A^2)}{2A^2(1+a^2A^2)(1-Ar_-)}. \ee

\vspace{0 mm}
\section{Rotating C-metric boost mass and first law} \label{sec:SCFirst}

The only remaining quantity to compute is the energy of the (patch of) spacetime itself. For this we turn to the covariant phase space formalism's generalized Noether theorem, which takes a generalized Killing vector, a codimension two surface, and a set of parameter variations (e.g. $\delta A, \delta m, \delta q, \delta a, \delta \alpha$), and gives the associated charge variation. For example if we take our generalized Killing vector to be an infinitesimal gauge transformation with constant gauge parameter and our surface to be a sphere enclosing the black hole, then the charge variation is simply $\delta Q$, the electric charge of the black hole. ``Charge variation" is somewhat of a misnomer, because it is not necessarily the total derivative of a well-defined charge. When it is, the charge variation is said to be integrable. We write generic charge variations with $\slashed{\delta}$ rather than $\delta$ to emphasize that they are not necessarily integrable. The exact formulas used are given in \cite{Ball:2020vzo}. See \cite{Compere:2006my} for an excellent review.

\begin{figure}
  \centering
  \begin{minipage}[b]{0.49\textwidth}
    \centering
        \includegraphics[width=\textwidth]{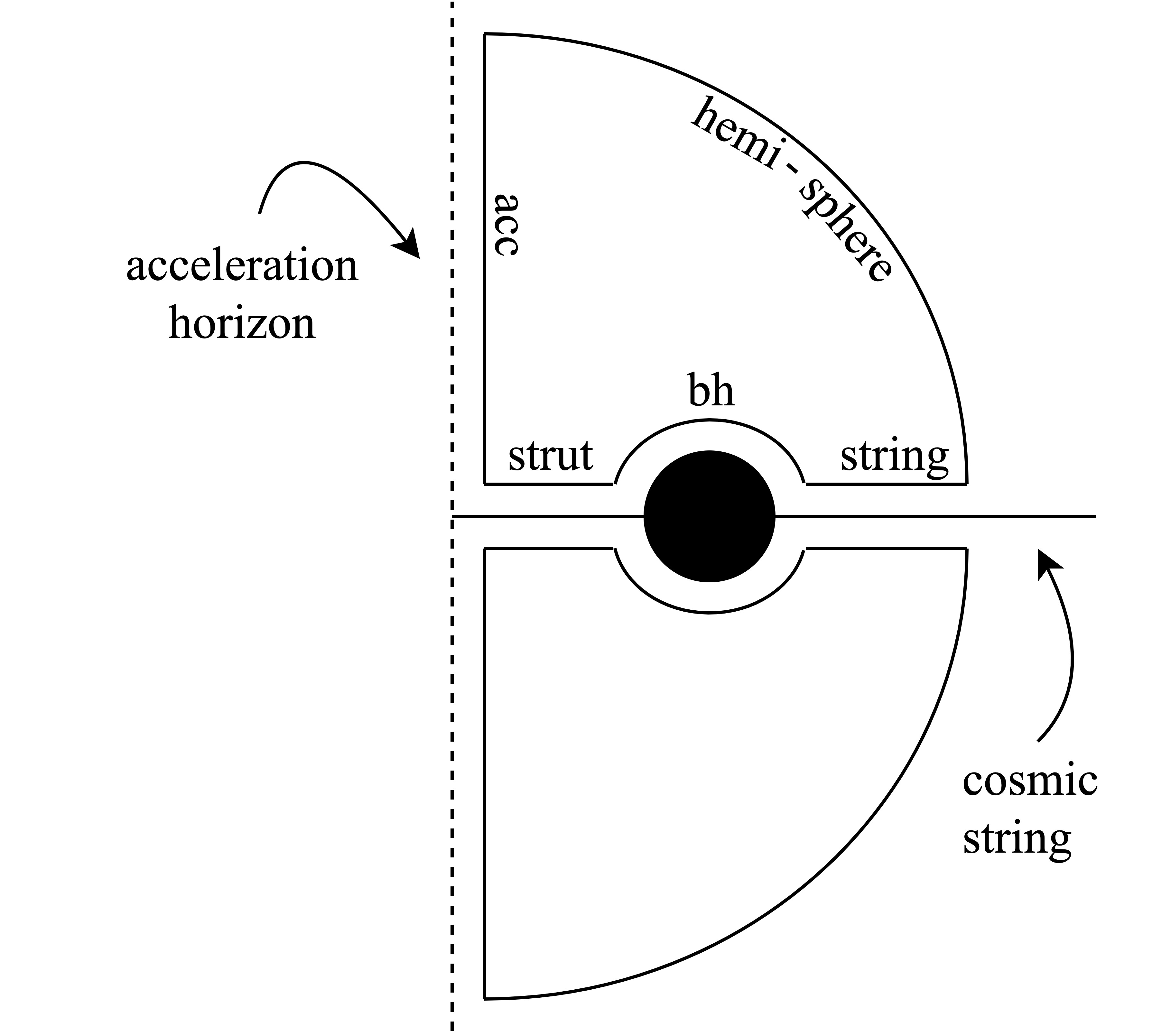}
    \caption{\label{fig:SCsurf} A contractible surface with pieces labelled, lying in a $t = \rm{const.}$ slice of the rotating C-metric. Shown with $\phi$ suppressed.}
  \end{minipage}
  \hfill
  \begin{minipage}[b]{0.49\textwidth}
  \centering
        \includegraphics[width=\textwidth]{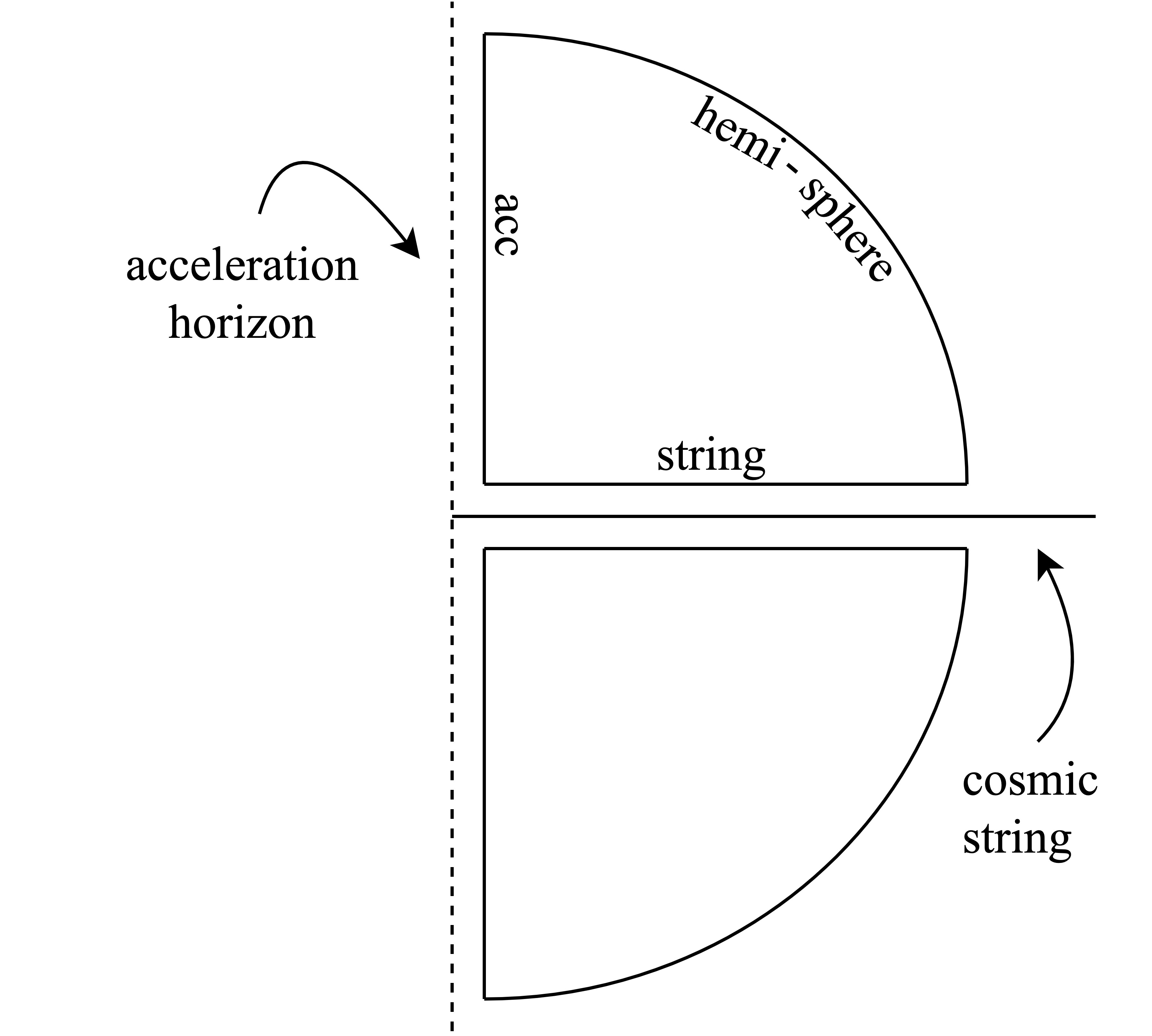}
    \caption{\label{fig:CSsurf} A contractible surface with pieces labelled, lying in a $t = \rm{const.}$ slice of the background. Shown with $\phi$ suppressed.}
  \end{minipage}
\end{figure}

We define the change in boost mass \cite{Dutta:2005iy} of our patch of spacetime relative to the background via the change in the charge variation associated with $\xi = N \p_t$ and a large hemisphere hugging spatial infinity. This is a natural generalization of the usual definition of the energy of a spacetime, which is the charge associated with the asymptotic sense of time, integrated over a large sphere near spatial infinity. One finds
\be \slashed{\delta} \Delta M = 0 \ee
where $\slashed{\delta}$ refers to the variation of parameters and $\Delta$ refers to the change relative to the background. Its vanishing is a consequence of the agreement of the rotating C-metric and cosmic string background metric, as well as their electromagnetic field strengths, at first subleading order in $\eps$. Zero is always integrable, so we conclude that $\Delta M$ is well-defined and constant on parameter space. Furthermore, this constant must be zero since our parameter space includes the cosmic string background itself. So we have
\be \Delta M = 0. \ee
The charge variation is invariant under deformations of the surface not passing through matter (i.e. the string). Following the approach of \cite{Ball:2020vzo}, we use this to equate $\delta \Delta M$ with the sum of several other charge variations,
\be \label{surfparts} \delta \Delta M = \slashed{\delta} \Delta \cq_{\rm acc} + \slashed{\delta} \cq_{\rm bh} + \slashed{\delta} \cq_{\rm strut} + \slashed{\delta} \Delta \cq_{\rm string}. \ee
These charge variations are all associated with $\xi$, but use different surfaces indicated by their subscripts. See figures \ref{fig:SCsurf} and \ref{fig:CSsurf}. In general if $\chi$ is the Killing generator of a horizon, and we take our surface to be a section of that horizon, then the associated charge variation is $\frac{\kap}{8\pi} \delta \ca - \Phi \delta Q$ (up to signs associated with orientation) where $\kap$ is $\chi$'s surface gravity, $\ca$ is the horizon area, $\Phi$ is the horizon potential, and $Q$ is the horizon electric charge \cite{Compere:2006my}. Thus by rewriting $\xi = \chi_{\rm bh} - \Omega_{\rm bh} \p_\phi = \chi_{\rm acc} - \Omega_{\rm acc} \p_\phi$ we can deduce
\be \ba \slashed{\delta} \cq_{\rm bh} & = \frac{\kap_{\rm bh}}{8\pi} \delta \ca_{\rm bh} - \Phi_{\rm bh} \delta Q - \Omega_{\rm bh} \delta J, \\
\slashed{\delta} \Delta \cq_{\rm acc} & = \frac{\kap_{\rm acc}}{8\pi} \delta \Delta \ca_{\rm acc} + \Phi_{\rm acc} \delta Q + \Omega_{\rm acc} \delta J. \ea \ee
This can also be checked directly. For the strut and string we find
\be \ba \slashed{\delta} \cq_{\rm strut} & = \ell_{\rm strut} \delta \mu_{\rm strut}, \\
\slashed{\delta} \Delta \cq_{\rm string} & = \Delta \ell_{\rm string} \delta \mu_{\rm string}. \ea \ee
Substituting our results back in \eqref{surfparts} gives the classical first law:
\be \label{SCfirst} \delta \Delta M = 0 = \frac{\kap_{\rm acc}}{8\pi} \delta \Delta \ca_{\rm acc} + \frac{\kap_{\rm bh}}{8\pi} \delta \ca_{\rm bh} + (\Phi_{\rm acc} - \Phi_{\rm bh}) \delta Q + (\Omega_{\rm acc} - \Omega_{\rm bh}) \delta J + \ell_{\rm strut} \delta \mu_{\rm strut} + \Delta \ell_{\rm string} \delta \mu_{\rm string}. \ee
All the quantities here have physical meaning independent of their roles in the formula. In particular, the thermodynamic lengths{\footnote{The use of ``thermodynamic length" here is unrelated to the use of the same term in \cite{Crooks}.}, i.e. the quantities multiplying the variations in strut/string tension} \cite{Appels:2017xoe}, are simply the generalized lengths of the strut and string \cite{Herdeiro:2009vd, Krtous:2019fpo}. The quantities can be interpreted as being seen by an asymptotic Rindler observer with proper acceleration $\kap_{\rm acc}$. We could also take the point of view of a global meta-observer and normalize $\xi$ through the asymptotic Killing algebra, which amounts to choosing $N$ such that $\kap_{\rm acc} = 1$. Of these two perspectives, the latter seems more relevant to celestial holography. Note that \eqref{SCfirst} holds independently of any choice of normalization, gauge potential, or rotation frame. The normalization $N$ in $\xi = N \p_t$ is arbitrary. A constant shift of the gauge potential would cancel between $\Phi_{\rm acc}$ and $\Phi_{\rm bh}$. Likewise, a change in the rotation frame would shift $\Omega_{\rm acc}$ and $\Omega_{\rm bh}$ by the same amount, ultimately cancelling. The thermodynamic interpretation of first laws such as \eqref{SCfirst} is discussed in \cite{Ball:2020vzo}.

With the first law in hand one can derive a Smarr-like relation by dimensional analysis \cite{acovErnst}. Explicitly,
\be 0 = \frac{\kap_{\rm acc}}{8\pi} \Delta \ca_{\rm acc} + \frac{\kap_{\rm bh}}{8\pi} \ca_{\rm bh} + \half (\Phi_{\rm acc} - \Phi_{\rm bh}) Q + (\Omega_{\rm acc} - \Omega_{\rm bh}) J. \ee
The strut and string terms have dropped out because tension is dimensionless in geometrized units.

\section{Ernst metric review} \label{sec:ErnstRev}

The similarities between the Ernst metric and the rotating C-metric are manifold, and we have attempted in our notation to make them as manifest as possible. The electric Ernst metric with conical deficits can be written as
\be \label{Ernst} ds^2 = \frac{\Lam(y,x)^2}{A^2 (x-y)^2} \left( G(y) dt^2 - \frac{dy^2}{G(y)} + \frac{dx^2}{G(x)} + \frac{G(x)}{\Lam(y,x)^4} \alpha^2 d\phi^2 \right). \ee
It is an exact solution of the Einstein-Maxwell-Nambu-Goto equations, with cosmic struts/strings along the $\phi$-axis and sole nonzero gauge field component
\be A_t = \frac{E \, G(y) (2+qE(2x-y))}{-4A^2(x-y)^2} + \frac{y}{4A^2} \left[ 2A(2-qEy)(qA - mE) + qE^2 (1 + A^2q^2(y^2 - 1)) \right]. \ee
The function $G(\zeta)$ is the quartic polynomial
\be G(\zeta) = (1 - \zeta^2)(1 + A r_+ \zeta)(1 + A r_- \zeta) \ee
with
\be r_\pm = m \pm \sqrt{m^2 - q^2}. \ee
As in the rotating C-metric, we assume $A r_+ < 1$. We also have the auxiliary function
\be \Lam(y,x) = \left( 1 + \half q E x \right)^2 + \frac{E^2 G(x)}{4A^2(x-y)^2}. \ee
We identify $\phi \sim \phi + 2\pi$ with other coordinates held fixed. The range of $x$ is $-1 \le x \le 1$. The $\phi$-axis between the black holes corresponds to $x=1$, while the outer $\phi$-axis corresponds to $x=-1$. We will concentrate on the patch $\frac{-1}{Ar_+} \le y \le -1$ which contains one of the two black holes, and in which $\p_t$ is a timelike Killing vector. The acceleration horizon lies at $y=-1$ and the black hole horizon lies at $y=\frac{-1}{Ar_+}$. If we continued past the first black hole horizon we would encounter its second, inner horizon at $y=\frac{-1}{Ar_-}$. Spatial infinity is reached by sending $y, x \to -1$ simultaneously with $\frac{1+x}{-1-y}$ fixed.

The Ernst metric and related spacetimes have been discussed in \cite{Ernst:1976mzr, Garfinkle:1993xk, hhrErnst, Gibbons:2013dna, acovErnst, AstReg}. Physically \eqref{Ernst} describes two charged black holes pulled apart to null infinity by a background electric field, helped or hindered by cosmic struts/strings. It is specified by five parameters, such as $A, m, q, E, \alpha$. At a loose intuitive level, the first four respectively give the black hole acceleration, the black hole mass, the black hole electric charge, and the strength of the background electric field. The parameter $\alpha$ could be absorbed into the range of $\phi$, but we keep $0 < \phi < 2\pi$. There is generically a strut at $x=1$ and a string at $x=-1$ on the $\phi$-axis, although authors often set $\alpha, A$ to eliminate them. Sending $E \to 0$ yields the charged C-metric, in the coordinates of \cite{Ball:2020vzo}. Alternatively, sending $m, q \to 0$ with $A, E, \alpha$ fixed yields the Melvin metric (discussed momentarily) in an accelerating coordinate system, generically with a cosmic string.

\section{Melvin background} \label{sec:Melv}

\begin{figure}
  \centering
  \begin{minipage}[b]{0.75\textwidth}
    \centering
        \includegraphics[width=\textwidth]{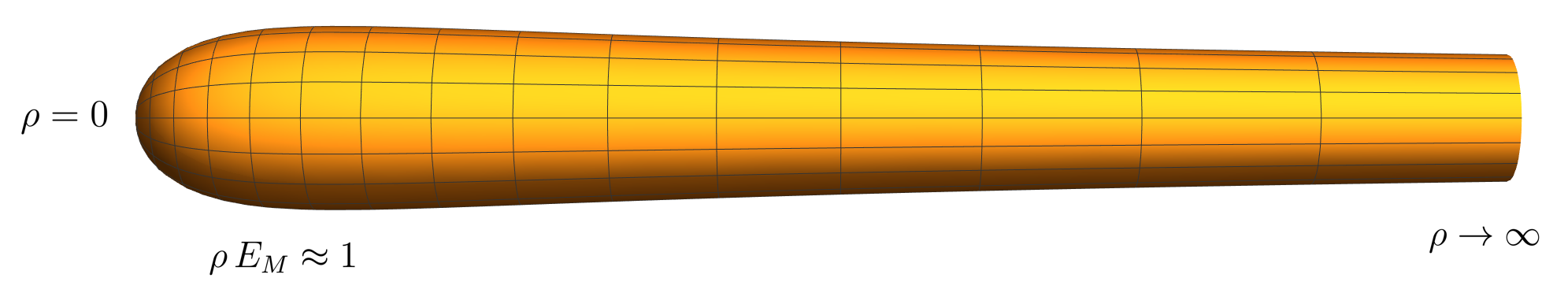}
    \caption{\label{fig:MelvSlice} A constant $T, Z$ slice of Melvin spacetime, embedded in $\mathbb{R}^3$. We can see that the circumference shrinks at large $\rho$.}
  \end{minipage}
  \hfill
\end{figure}

We would like to find a suitable background for the Ernst metric, like we did for the rotating C-metric. The natural background for the electric Ernst metric (with cosmic string) is the electric Melvin metric \cite{Melvin:1963qx} (with cosmic string), which can be written
\be \label{Melvin} ds^2 = \left( 1 + \frac{1}{4} E_M^2 \rho^2 \right)^2 \left[ -dT^2 + dZ^2 + d\rho^2 \right] + \frac{\rho^2 \alpha_M^2 d\phi^2}{\left( 1 + \frac{1}{4} E_M^2 \rho^2 \right)^2}, \ee
with gauge field
\be A_\mu dx^\mu = Z E_M dT. \ee
The electric field dies off at $E_M \rho \gg 1$, and along the $\phi$-axis has magnitude $E_M$:
\be \half F_{\mu\nu} F^{\mu\nu} \Big|_{\rho = 0} = -E_M^2. \ee
There is also a cosmic string with deficit angle $\delta_M = 2\pi (1 - \alpha_M)$ on the $\phi$-axis. Physically the Melvin metric describes a cylindrically symmetric spacetime with a ``tube" of electric flux concentrated near the $\phi$-axis, held together by the gravitational attraction of its own energy. The total amount of flux is finite, equal to the amount of flux sourced by a charge
\be \label{Melvcharge} Q_M \equiv -\frac{\alpha_M}{E_M}. \ee
The Melvin metric is not asymptotically locally flat. Its asymptotics can be understood by considering a surface of constant $T$ and $Z$ in \eqref{Melvin}, on which the induced metric is
\be ds^2_{\rm slice} = \left( 1 + \frac{1}{4} E_M^2 \rho^2 \right)^2 d\rho^2 + \frac{\rho^2 \alpha_M^2 d\phi^2}{(1 + \frac{1}{4} E_M^2 \rho^2)^2}. \ee
The embedding of this surface in $\mathbb{R}^3$ is shown in figure \ref{fig:MelvSlice} for $\alpha_M = 1$. The radial direction extends infinitely, but at large radius the circumference shrinks. Near $\rho = 0$ the slice is approximately a cone (or just flat when $\alpha_M = 1$).

Spatial infinity of Ernst is more easily described in coordinates $(t, \eps, \chi, \phi)$, defined by substituting
\be \ba y & = -1 - \eps (1-\chi) \\
x & = -1 + \eps \chi \ea \ee
in \eqref{Ernst}. Now spatial infinity is reached by $\eps \to 0$, with $\chi, \phi$ parametrizing the direction. Following the approach of \cite{hhrErnst}, the Melvin metric can be written in $(t,\eps,\chi,\phi)$ coordinates such that asymptotically its components differ from the Ernst metric only at sub-subleading order in $\eps$. This story is identical in spirit to that of the rotating C-metric. The Melvin background corresponding to our Ernst metric has
\be E_M = \frac{E (1 - A r_+)(1 - A r_-)}{(1 - \half q E)^3} \ee
and
\be \alpha_M = \frac{G'(-1)}{2(1-\half qE)^4} \alpha. \ee
The Ernst and Melvin gauge fields can be chosen to agree exactly on the acceleration horizon, but they will still differ at first subleading order in $\eps$ elsewhere. Combined with the persistence of the axial field strength out to infinity, this is enough to give a nonzero change in boost mass variation.

\section{Ernst metric quantities and first law} \label{sec:Ernstqnf}

Here we record some of the Ernst metric's physical quantities in our conventions. All can be viewed as functions of $A, m, q, E, \alpha$. The black hole's area is
\be \ca_{\rm bh} = \frac{4\pi \alpha r_+^2}{1 - A^2 r_+^2}. \ee
The change in acceleration horizon area compared to the Melvin background is
\be \Delta \ca_{\rm acc} = -\frac{4\pi\alpha (m - Aq^2)}{A(1 - A r_+)(1 - A r_-)}. \ee
The electric charge of the black hole is
\be Q = \frac{1}{4\pi} \oint dx d\phi \sqrt{-g} F^{yt} \big|_{y=\frac{-1}{Ar_+}} = \frac{\alpha q}{1 - \frac{1}{4} q^2 E^2}. \ee
The electric charge associated with the acceleration horizon is
\be Q_{\rm acc} = \frac{1}{4\pi} \int dx d\phi \sqrt{-g} F^{yt} \big|_{y=-1} = -\frac{\alpha}{E (1 + \half q E)}. \ee
For the change relative to the background charge \eqref{Melvcharge} one finds
\be \Delta Q_{\rm acc} \equiv Q_{\rm acc} - Q_M = \frac{\alpha q}{1 - \frac{1}{4} q^2 E^2}. \ee
Note this is just $Q$. The strut and string tensions are
\be \ba \mu_{\rm strut} & = \frac{1}{4} \left( 1 - \frac{\alpha |G'(1)|}{2(1 + \half q E)^4} \right), \\
\mu_{\rm string} & = \frac{1}{4} \left( 1 - \frac{\alpha \, G'(-1)}{2(1 - \half q E)^4} \right). \ea \ee
Several of our quantities of interest are defined in terms of a timelike Killing vector. We will use $\xi = N \p_t$ throughout, with normalization $N(A, m, q, E, \alpha)$. The surface gravity of the black hole horizon with respect to $\xi$ is
\be \kap_{\rm bh} = \half N \big| G'(\frac{-1}{A r_+}) \big|. \ee
The surface gravity of the acceleration horizon with respect to $\xi$ is
\be \kap_{\rm acc} = \half N G'(-1). \ee
The electric potential at the black hole horizon is
\be \Phi_{\rm bh} \equiv (\xi^\mu A_\mu)|_{y=\frac{-1}{A r_+}} = N \frac{2Er_+ - 4Aq + Aq^3 E^2}{4A^2 r_+}. \ee
The electric potential at the acceleration horizon is
\be \Phi_{\rm acc} \equiv (\xi^\mu A_\mu)|_{y=-1} = -N \frac{(qE^2 + 2A(2+qE) (Aq - mE))}{4A^2}. \ee
The generalized lengths of the strut and string are
\be \ell_{\rm strut} = N \frac{(1-A r_+) (1 + \half q E)^4}{2A^2 (1 + A r_+)} \ee
and
\be \Delta \ell_{\rm string} = -N \frac{(1 - 3Ar_-) (1 - \half q E)^4}{2A^2 (1 - A r_-)}. \ee
The last quantity is the variation in the boost mass of the (patch of) spacetime. We compute it with covariant phase space methods as the change between the Ernst metric and Melvin background of the charge variation associated with $\xi$ and a large surface near spatial infinity. In this case the boost mass variation does not vanish. One finds
\be \slashed{\delta} \Delta M = P \, \delta \left( \frac{E_M}{1 - 4\mu_{\rm string}} \right), \ee
where $\delta (\dots)$ represents the variation with respect to $A, m, q, E, \alpha$ and we have defined
\be P \equiv N \frac{G'(-1) \alpha^2 q}{4A^2 (1 - \half q E)^2}. \ee
We call $P$ the electric dipole moment of the Ernst metric because it is essentially the ratio of the energy variation to the external field strength variation. The factor $1-4\mu_{\rm string}$ is precisely $\frac{1}{2\pi}$ times the ratio of circumference to radius around a flat cosmic string of tension $\mu_{\rm string}$. Our dipole moment $P$ is unrelated to those of \cite{Gibbons:2013dna, acovErnst}, which were defined by varying the energy of the black hole as opposed to the energy of the spacetime itself. At least na\"ively, it is highly nontrivial that $\slashed{\delta}\Delta M$ turns out to be proportional to a total derivative in our five-parameter space. We have not made any normalization or gauge choices to contrive this result. One might use the normalization freedom to set $P = 1$, in which case the boost mass is integrable. This is a valid choice, but the presence of energy flux from infinity means one should not necessarily expect integrability of the boost mass, and we leave the normalization general in what follows.

Although the shape of spacetime here is different from the rotating C-metric case, the topological aspects of the argument equating surface charges to derive the first law are entirely unchanged. The classical first law for Ernst reads
\be \label{fErnst} P \, \delta \left( \frac{E_{M}}{1-4\mu_{\rm string}} \right) = \frac{\kap_{\rm acc}}{8\pi} \delta \Delta \ca_{\rm acc} + \frac{\kap_{\rm bh}}{8\pi} \delta \ca_{\rm bh} + (\Phi_{\rm acc} - \Phi_{\rm bh}) \delta Q + \ell_{\rm strut} \delta \mu_{\rm strut} + \Delta \ell_{\rm string} \delta \mu_{\rm string}. \ee
Note this holds independently of any choice of normalization or gauge potential. The quantities can be interpreted as those seen by a static observer near spatial infinity, of proper acceleration $\kap_{\rm acc}$. See \cite{Ball:2020vzo} for a discussion of the thermodynamic interpretation of first laws like \eqref{fErnst}.

Once we have the first law we can derive a Smarr-like relation from it using dimensional analysis. One finds
\be -\half P \frac{E_M}{1 - 4\mu_{\rm string}} = \frac{\kap_{\rm acc}}{8\pi} \Delta \ca_{\rm acc} + \frac{\kap_{\rm bh}}{8\pi} \ca_{\rm bh} + \half (\Phi_{\rm acc} - \Phi_{\rm bh}) Q. \ee
The strut and string terms have dropped out because tension is dimensionless in our units.

\section{Conclusion} \label{sec:conc}

We have seen that the rotating C-metric and Ernst metric both describe accelerating black holes, with the acceleration provided by a combination of struts, strings, and/or electromagnetic force. They are direct generalizations of both Rindler space and Reissner-Nordstr\"om black holes. The flat cosmic string spacetime naturally served as a background for the rotating C-metric, while the Melvin spacetime did so for the Ernst metric. Some quantities, like the electric charge, had obvious definitions, but others, like the generalized length of the string, were more subtle. Divergent quantities like the area of the acceleration horizon were regularized by carefully comparing with the respective background. Covariant phase space methods were used to define the boost mass, which was zero for the rotating C-metric but nonzero for the Ernst metric, motivating the definition of a dipole moment. With all the pieces in place we saw that the rotating, charged C-metric and the Ernst metric each satisfied a first law directly generalizing that of \cite{Ball:2020vzo}. These results emphasize the robustness of the physical principles underlying the more familiar Kerr-Newman first law and pave the way for further explorations of accelerating black hole thermodynamics in flat space, which had been stalled for some time due to uncertainty about the incorporation of the acceleration horizon and external force.

One of the advantages of the approach to first laws used in this paper is that the clear physical interpretations and mathematical definitions of the quantities make generalization relatively straightforward. For example the expression for $\Delta \ell_{\rm string}$ for the rotating C-metric was more complicated than it was for the C-metric, but it still had the same physical interpretation, namely as the change in generalized length of the string. So while every term in the rotating C-metric first law was different in detail from the C-metric first law, most had clear definitions from the start. The main conceptual novelties were the angular momentum term and the mixing of $\p_t$ and $\p_\phi$. And even then, the covariant phase space formalism was happy to show us the way. For these reasons we are optimistic about the generalization of our results to larger classes of metrics. An obvious guess for the first law of the rotating Ernst metric is to simply combine \eqref{SCfirst} and \eqref{fErnst}. It also seems likely that the accelerating multi-black hole spacetime of \cite{Dowker:2001dg} will simply have an area term for each black hole and a tension term for each string. Another direction for generalization is to the entire seven-parameter Plebanski-Demianski family of boost-rotation symmetric solutions \cite{Plebanski:1976gy}, although there may be some subtleties in handling NUT charge.

There is also further work to do on the thermodynamic interpretation of our first laws. It was shown in \cite{Ball:2020vzo} that the na\"ive grand potential associated with the C-metric first law does agree with the semiclassical partition function, but an understanding of how to derive the thermodynamic quantities through differentiation of a thermodynamic potential is still lacking.
In any case, these fascinating and intricate exact solutions contain an inordinate amount of physics, hinted at in the introduction, and we hope that our results can facilitate the continued unraveling of their mysteries.

\acknowledgments

We are grateful to Noah Miller for collaboration at an early stage of the work and to Andy Strominger for useful discussions. We would also like to thank Ji\v{r}\'i Podolsk\'y for emphasizing to us the physical inequivalence of rotating C-metrics using the two different common forms of the polynomial $G(\zeta)$.

We gratefully acknowledge support from NSF grant 1707938. This work is funded in part by the Gordon and Betty Moore Foundation. It was also made possible through the support of a grant from the John Templeton Foundation. The opinions expressed in this work are those of the author and do not necessarily reflect the views of the John Templeton Foundation.

\bibliography{cthermo}
\bibliographystyle{utphys}

\end{document}